\documentclass[preprint,pre,aps,superscriptaddress,showpacs]{revtex4}

\usepackage{amsmath}
\usepackage{graphicx}

\newcommand{\beq}{\begin{equation}}
\newcommand{\eeq}{\end{equation}}
\newcommand{\be}{\begin{equation}}
\newcommand{\ee}{\end{equation}}
\newcommand{\bea}{\begin{eqnarray}}
\newcommand{\eea}{\end{eqnarray}}
\newcommand{\barr}{\begin{array}}
\newcommand{\earr}{\end{array}}

\begin{document}

\title{Driven low density granular mixtures.}

\author{Riccardo Pagnani}
\affiliation{
Dipartimento di Fisica \\
Universit\'a ``La Sapienza'',P.le A. Moro 2, 00198 Roma, Italy}
\affiliation{
Istituto Nazionale di Fisica della Materia, Unit\`a di Roma, Roma, Italy}
\author{Umberto \surname{Marini Bettolo Marconi}}
\affiliation{
Dipartimento di Fisica \\
Universit\'a di Camerino, 62032 Camerino, Italy}
\affiliation{
Istituto Nazionale di Fisica della Materia, Unit\`a di Camerino, Camerino, Italy}
\author{Andrea Puglisi} 
\affiliation{
Dipartimento di Fisica \\
Universit\'a ``La Sapienza'',P.le A. Moro 2, 00198 Roma, Italy}
\affiliation{
Istituto Nazionale di Fisica della Materia, Unit\`a di Roma, Roma, Italy}

\date{\today}

\begin{abstract}
We study the steady state properties of a 2D granular mixture in the
presence of energy driving by employing simple analytical estimates
and Direct Simulation Monte Carlo. We adopt two different driving
mechanisms: a) a homogeneous heat bath with friction and b)
a vibrating boundary (thermal or harmonic) in the presence 
of gravity. The main findings
are: the appearance of two different granular temperatures, one for each
species; the existence of overpopulated tails in the velocity
distribution functions  and of non trivial spatial correlations
indicating the spontaneous formation of cluster aggregates. In the
case of a fluid subject to gravity and to a vibrating boundary, both
densities and temperatures display non uniform profiles along the
direction normal to the wall, in particular the temperature profiles
are different for the two species while the temperature ratio is
almost constant with the height. Finally, we obtained the velocity
distributions at different heights and verified the non gaussianity of
the resulting distributions.

\end{abstract}
       
\maketitle


\section{Introduction}

Granular materials present a rich and intriguing phenomenology, which
has attracted the interest of the scientific community since the
nineteenth century \cite{general}. However, in spite of its recent progress the
theoretical study of granular gases, i.e. of fluidized granular
particles, is certainly less advanced than that concerning ordinary
molecular fluids. The reason for this state of affairs is the presence
of dissipation due to inelastic collisions and of friction with the
surroundings, which prevents these system to reach thermodynamic
equilibrium. In fact, in order to render stationary a granular system
one needs to inject continuously energy into the system. This can be
done, for instance by shaking or vibrating the grains.

In the present paper we illustrate the results of a numerical
investigation concerning the properties of a two component granular
mixture, modeled, following an established tradition, by inelastic hard
spheres (IHS) with different masses, restitution coefficients, radii,
and subject to different forms of external drive. The physical motivation
for our study stems from the fact that in nature most granular
materials are polydisperse from the point of view of their sizes
and/or of their physical and mechanical properties. The theoretical
study of granular mixtures has attracted so far the attention of several
researchers\cite{dufty,ioepuglio,Barrat,duftynew}. 
These studies comprehend both freely cooling and  uniformly
heated granular mixtures and have been performed almost contemporaneously
with laboratory experiments \cite{feitosa,Parker}.
The most striking outcome is the lack of energy equipartition,
i.e. the presence of two different kinetic temperatures, one for
each species.

In the present paper our interest will be concentrated on the
stationary state obtained by applying an energy feeding mechanism,
represented by a uniform stochastic driving force or by a vibrating
wall.  In the first case the
work of Barrat and Trizac \cite{Barrat} and Garz\'o et al. \cite{duftynew} 
provides a simple theoretical account, based
on the Boltzmann equation approach and on 
the Direct Simulation Monte Carlo 
method (DSMC) \cite{bird1,bird2}, of the lack of energy
equipartition.

The present work differs from the recently appeared papers dealing with
granular mixtures in several aspects:

\begin{itemize} 
\item
the heat bath, which drives the system, is
at finite temperature, since we have introduced a finite friction of
the particles with the surroundings and we have considered the possibility
of having spatially inhomogeneous states;

\item
in addition, we considered a second situation in which the particles
subject to a vertical gravitational field receive energy
anisotropically from the bottom vibrating wall of the container;

\item
in both cases we provide information about the presence of
inhomogeneities in the system, e.g. density clustering and non-uniform
density and temperature profiles.

\end{itemize} 

In Sec. II we present the model fluid and two different  
mechanisms of energy supply. In sect III we discuss the first sub-model, the
one with the heat bath, and obtain by means of an approximate analytic
method an estimate of the partial temperature of each component.
Subsequently we study the same sub-model with a DSMC algorithm. In
sect. IV we study numerically the second sub-model (the one 
with gravity and vibrating
wall) by means of DSMC simulations.  Finally in the last section, V,
we present our conclusions.


\section{Definition of the models}

We shall consider a dilute inelastic gas constituted of $N_1$
particles of mass $m_1$ and $N_2$ particles of mass $m_2$ subject to
some kind of external driving (this will be specified in the following).
 We suppose that the interactions between the grains can be
described by the smooth inelastic hard sphere model (IHS)
\cite{campbell}, thus we specify only the radius of the
spheres, their masses and the fraction of the kinetic energy
dissipated at each collision. This can be done by defining three
different restitution coefficients $\alpha_{ij}$, i.e. $\alpha_{11}, \alpha_{22}$,
and $\alpha_{12}=\alpha_{21}$, which account for normal dissipation in collisions
among particles of type $i$ and $j$. No internal degrees of freedom
(e.g. rotations) are included.

One can describe the velocity changes induced by the instantaneous
inelastic collisions of smooth disks labeled $1$ and $2$ of diameter
$\sigma_1$ and $\sigma_2$ by the following equations:

\begin{subequations}
\begin{align}
{\bf v}_1' &={\bf v}_1-\frac{1+\alpha_{\kappa_1\kappa_2}}{2}
\frac{m_{\kappa_2}}{m_{\kappa_1}+m_{\kappa_2}}
(({\bf v}_1-{\bf v}_2) \cdot {\bf
\hat{n}}){\bf \hat{n}}  \\
{\bf v}_2' &={\bf v}_2+\frac{1+\alpha_{\kappa_1\kappa_2}}{2}
\frac{m_{\kappa_1}}{m_{\kappa_1}+m_{\kappa_2}}
(({\bf v}_1-{\bf v}_2) \cdot {\bf
\hat{n}}){\bf \hat{n}} 
\end{align}
\label{collision}
\end{subequations}

where ${\bf \hat{n}}=2({\bf x_1}-{\bf
x_2})/(\sigma_{\kappa_1}+\sigma_{\kappa_2})$ is the unit vector along the
line of centers ${\bf x}_1$ and ${\bf x}_2$ of the colliding disks at
contact and $\kappa_1, \kappa_2$ are the species ($1$ or $2$) to whom
particles $1$ and $2$ belong.  An elementary collision
conserves the total momentum and reduces the relative kinetic energy
by an amount proportional to $(1-\alpha_{\kappa_1\kappa_2}^2)/4$.
The collision rule we have adopted excludes the presence of tangential
forces, and hence the rotational degrees of freedom do not contribute
to the description of the dynamics.

Since the particles suffer mutual collisions and loose kinetic energy,
in order to achieve a steady state, one needs to supply from the
exterior some energy. The energy source has been modeled in two
different fashions.

In the first sub-model we have assumed that the particles, besides
suffering mutual collisions experience a 
uniform stochastic force and a
viscous damping\cite{puglisi1,puglisi2}.  
The presence of the frictional, velocity-dependent
term in addition to the random forcing, not only is
motivated by the idea of preventing the energy of a driven elastic
system ($\alpha_{\kappa_1\kappa_2} \to 1$), to increase indefinitely, but also
mimics the presence of friction of the particles with the
container. Moreover a fluctuation dissipation relation is assumed
between the viscous force and the intensity of the noise.  Even in
extended systems with small inelasticity the absence of friction may
cause some problems of stability~\cite{tobepubl}.  
We shall be mainly interested in
the stationary situation determined by the balance between the energy
feeding mechanism and the dissipative forces due both to the
inelasticity of the collisions and to the velocity-dependent frictional
force.

In the second
sub-model the grains are
constrained to move on a frictionless inclined plane and the bottom boundary
vibrates (as a thermal~\cite{japan} or deterministic~\cite{kudrolli2}
oscillating wall) giving energy only to the particles bouncing on it.
Periodic boundary conditions are assumed laterally.

Since we consider throughout only sufficiently low density systems 
successive binary collisions are effectively uncorrelated and 
Boltzmann equation
can be used to describe the non equilibrium dynamics. In fact, by
assuming the validity of the Boltzmann molecular chaos hypothesis,
introduced to treat the collision term, it is straightforward to
derive the governing equations for the probability density
distribution function of each species.  Moreover in this situation we
are allowed to perform Direct Simulation Monte Carlo to 
integrate numerically the inhomogeneous Boltzmann
equation of the sub-models.



\section{Uniformly heated system}

In order to see the effect of the heat bath let us consider the system
in the absence of collisions. In this case, the evolution of the
velocity of each particle is described by an Ornstein-Uhlenbeck
process.  If we require that the two components must reach the same
granular temperature in the limit of vanishing inelasticity we have
two different possibilities to fix the heat bath parameters:

\begin{equation}
\label{posit}
\partial_t \mathbf{x}_i(t) = \mathbf{v}_i(t)
\end{equation}

\begin{subequations}
\label{riclangevin}
\begin{align}
m_i \partial_t \mathbf{v}_i(t)&=-\gamma \mathbf{v}_i(t)+\sqrt{2 \gamma T_b}\;\xi_i(t) \label{recipe1} \\
m_i \partial_t \mathbf{v}_i(t)&=- m_i \beta \mathbf{v}_i(t)+\sqrt{2 m_i \beta T_b}\;\xi_i(t) \label{recipe2}
\end{align}
\end{subequations}
where $i=1,2$ and $T_b$ is the heat bath temperature
and $\xi(t)$ is a Gaussian noise with the following properties:
\begin{subequations}
\begin{align}
<\xi_i(t)>&=0 \\
<\xi_i(t_1) \xi_j(t_2)>&=\delta(t_1-t_2)\delta_{ij}
\end{align}
\end{subequations}

The associated Fokker-Planck equations for the two cases are
respectively:

\begin{subequations}
\begin{align}
\partial_t f_i(\mathbf{r,v},t)&=
\frac{\gamma}{m_i}\nabla_v (\mathbf{v} f_i(\mathbf{r,v},t))+ \frac{\gamma T_b}{m_i^2}
\nabla^2_{v} f_i(\mathbf{r,v},t)+
\mathbf{v} \nabla_r f_i(\mathbf{r,v},t)  \\
\partial_t f_i(\mathbf{r,v},t)&=
\beta \nabla_v (\mathbf{v} f_i(\mathbf{r,v},t))
+\frac{\beta T_b}{m_i } \; \nabla^2_v f_i(\mathbf{r,v},t)+
\mathbf{v} \nabla_r f_i(\mathbf{r,v},t)
\end{align}
\label{bagni}
\end{subequations}


\subsection{Spatially Uniform solutions}

When we take into account collisions among particles 
equations~\eqref{bagni} become two coupled Boltzmann
equations modified by the presence of a diffusion term due to the 
thermal noise.
 We shall first consider a spatially
homogeneous case in order to derive the temperature of each species in
the homogeneous stationary state. This represents a sort of mean field
approximation. In fact, in the original Boltzmann equations the collisions
are considered to occur only between particles at the same spatial
location, whereas here these can occur between arbitrary pairs of
particles regardless their spatial separation.  In the present
subsection we compute the second moments, $<v_i^2>$, of the
distribution functions in order to determine the partial temperatures
and their ratio. Although the method of derivation of the equations
for the partial temperatures is not original, we present it in order
to render the paper self-contained and because it shows the
differences between the particular heat bath we employed and those
chosen by other authors~\cite{Barrat}.  First, indicating by
$n_i=N_i/V$ the partial density of species $i$, we notice that both
eqs.~\eqref{bagni} possess the same equilibrium solution:

\begin{equation}
f_i(\mathbf{v})=n_i\left(\frac{m_i}{2 \pi T_b}\right)^{\frac{d}{2}}
e^{-\frac{m_i v^2}{2 T_b}}
\end{equation}
although the relaxation properties are different. Only upon adding the 
inelastic collision term the two species display different temperatures.
The resulting Boltzmann equation  for a granular mixture is
\cite{dufty,Barrat,ioepuglio}:

\begin{equation}
\partial_t f_i(\mathbf{v}_1;t)=\sum_{j} J_{ij}[\mathbf{v}_1|f_i,f_j]
+\frac{\xi_{0i}^2}{2}\nabla^2_v f_i+\beta_i\nabla_v\cdot(\mathbf{v}_1 f_i)
\end{equation}
where we have used a compact notation to represent the two different 
choices of heat bath:

\begin{itemize}
\item {\bf Case 1}
 
\begin{equation}
\begin{array}{r}
\xi_{0i}^2 \rightarrow \frac{2 \gamma T_b}{m_i^2} \\
\beta_i\rightarrow \frac{\gamma}{m_i}
\end{array}
\end{equation}

\item{\bf Case 2}
 
\begin{equation}
\begin{array}{c}
\xi_{0i}^2 \rightarrow \frac{2 \beta T_b}{m_i}\\
\beta_i\rightarrow \beta 
\end{array}
\end{equation}

\end{itemize}
and $J_{ij}[v_1|f_i,f_j]$ is the collision integral: 

\begin{equation} 
J_{ij}[v_1|f_i,f_j]\equiv \sigma_{ij}^2 \int
d\mathbf{v}_2
\int d\mathbf{\hat{\sigma}} \Theta(\hat{\sigma}  \cdot
\mathbf{g}_{12}) (\hat{\sigma} \cdot \mathbf{g}_{12})
[\alpha_{ij}^{-2}
f_i(\mathbf{v_1^\prime})f_j(\mathbf{v_2^\prime})-f_i(\mathbf{v_1})f_j(\mathbf{v_2})]
\end{equation}
The primed velocities are pre-collisional states, which can be
obtained by inverting eqs. (\ref{collision})

Due to the presence of the heat bath terms the system reaches
asymptotically a steady state, characterized by time independent
pdf's. By imposing the vanishing of the time derivatives we obtain:

\begin{equation} 
\sum_{j} J_{ij}[\mathbf{v}_1|f_i,f_j]+\frac{\xi_{0i}^2}{2}\nabla^2_v
f_i+\beta_i\nabla_v \cdot(\mathbf{v}_1 f_i)=0 
\end{equation}
or after integrating over $\mathbf{v_1}$:
\begin{equation} 
\sum_{j} \int d\mathbf{v}_1 v_1^2 J_{ij}[\mathbf{v}_1|f_i,f_j]+ 
\frac{\xi_{0i}^2}{2} \int d\mathbf{v}_1 v_1^2 \nabla^2_v f_i+ 
\beta_i\int d\mathbf{v}_1 v_1^2\nabla_v \cdot(\mathbf{v}_1 f_i)=0
\end{equation}
After simplifying the second and the third integral by integration by parts
and using the normalization property $\int f_i dv_i =n_i$  we find:

\begin{equation} 
\sum_{j} \int d\mathbf{v_1} v_1^2 J_{ij}[\mathbf{v}_1|f_i,f_j]
+n_i d \xi_{0i}^2 - 2 \beta_i\int d\mathbf{v}_1 v_1^2 f_i(\mathbf{v}_1)=0
\label{momento}
\end{equation}
By defining the partial temperature:
\begin{equation}
n_i T_i \equiv \frac{1}{d} \int d\mathbf{v_1} m_i v_1^2 f_i
\end{equation}
we can rewrite eq. \eqref{momento} in terms of the temperatures :
\begin{equation} 
T_i=\frac{m_i}{2d\beta_i}\left(\frac{1}{n_i} \sum_{j} \int d\mathbf{v_1}
v_1^2 J_{ij}[v_1|f_i,f_j]+d \xi_{0i}^2 \right) 
\label{tpart}
\end{equation}
Eq. \eqref{tpart} determines the partial temperatures once the $f_i$ are 
known. In practice one can obtain an estimate of $T_i$ by substituting 
two Maxwell distributions: 

\begin{equation}
f_i(v) = n_i\left(\frac{m_i}{2 \pi T_i}\right)^{\frac{d}{2}}
e^{-\frac{m_i v_1^2}{2 T_i}}
\nonumber
\end{equation}
and performing the remaining integrals (see \cite{dufty,Barrat})
one gets:

\begin{equation}
\begin{split}
&\frac{d \Gamma(d/2)}{m_i \pi^{(d-1)/2}}2\beta_i (T_b -  T_i) =
\sigma_{ii}^{d-1} n_i \frac{2(1-\alpha_{ii}^2)}{m_i^{3/2}} T_i^{3/2}\\
&+ \sigma_{ij}^{d-1} n_j \mu_{ji} \left[ \mu_{ji} (1-\alpha_{ij}^2)
\left( \frac{2T_i}{m_i} +  \frac{2T_j}{m_j} \right)
+  4 (1+\alpha_{ij})\frac{T_i - T_j}{m_1+m_2} \right]
\left( \frac{2T_i}{m_i} +  \frac{2T_j}{m_j} \right)^{1/2} 
\label{eq:T1T2}
\end{split}
\end{equation}
where $\mu_{ij}=m_i/(m_i+m_j)$.

By solving numerically the nonlinear  system of eqs. 
(\ref{eq:T1T2}) one obtains the
steady values of the 
partial temperatures in the spatially homogeneous situation.

\begin{figure}[htbp]
\begin{center}
\includegraphics[clip=true,width=0.7\textwidth]{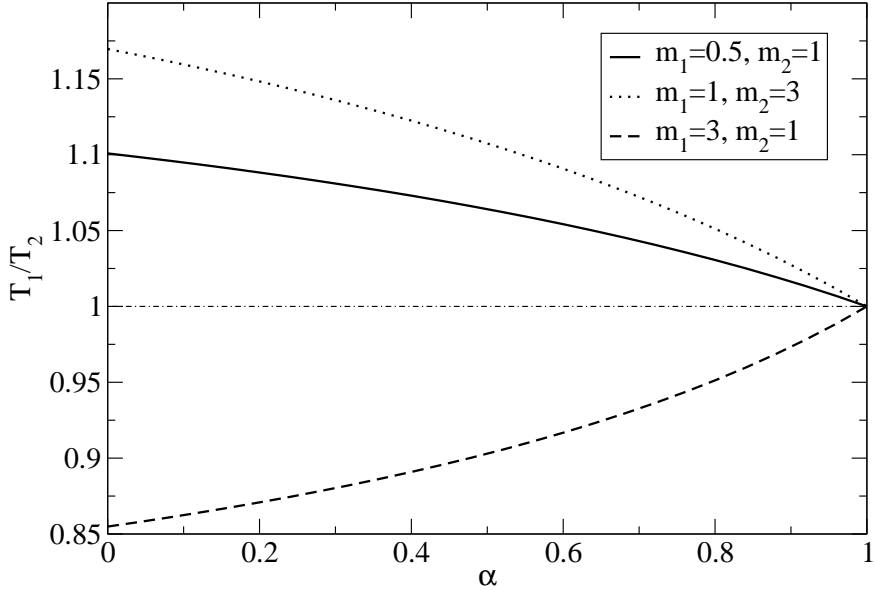}
\end{center}
\caption{
Granular temperature ratio $T_1/T_2$ vs. $\alpha$ obtained
with the heat bath, case 1, homogeneous, using $T_b=1$, 
$\gamma=0.1$ and different mass ratios.}
\label{simomotb1gamma01cicli005}
\end{figure}

\begin{figure}[htbp]
\begin{center}
\includegraphics[clip=true,width=0.7\textwidth]{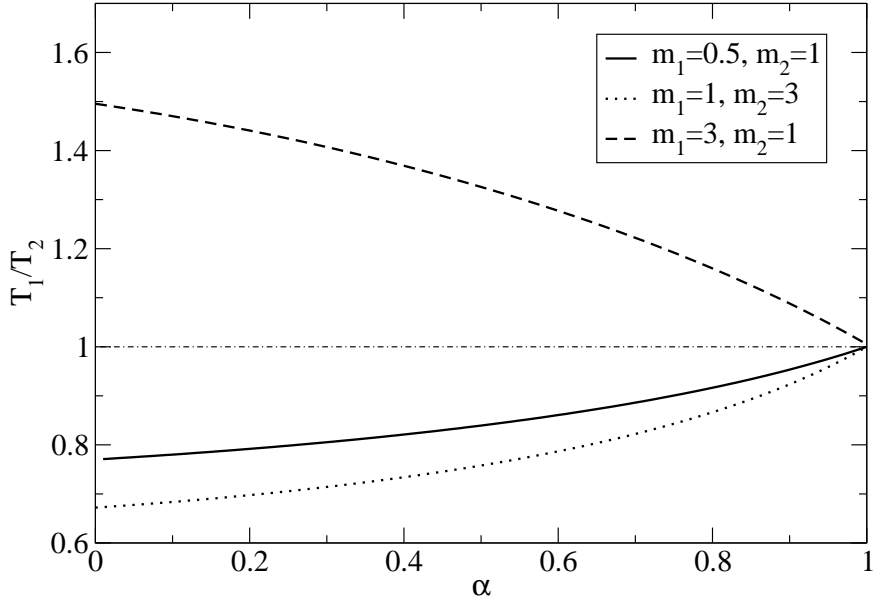}
\end{center}
\caption{
Granular temperature ratio $T_1/T_2$ vs. $\alpha$ using the second recipe
for the heat bath, case 2, homogeneous, with $T_b=1$, 
$\beta=0.1$ and various mass ratios.}
\label{ricetta2rapportotemperatura}
\end{figure}

In figs.~\ref{simomotb1gamma01cicli005}
and~\ref{ricetta2rapportotemperatura} we report the temperature ratio
$\frac{T_1}{T_2}$ as a function of 
a common restitution coefficient $\alpha$,
having chosen equal coefficients $\alpha_{11}=\alpha_{22}=\alpha_{12}=
\alpha$. Assuming  identical
concentrations, and a varying the mass ratio $\frac{m_1}{m_2}$, 
we studied the cases 1 and 2.

We notice that with the second recipe (case 2) the temperature ratio
is an increasing function of the mass ratio $m_1/m_2$: this is exactly
the opposite of what happens for the case 1. The experimental
observation~\cite{feitosa} suggest that the trend of case 2 is the
right one. 
In other words, in the case 2 both the friction term and the power 
supplied are proportional to the mass of the two species and
such a property is probably true also in the experiments.  
This is why we shall use the recipe of case 2 in the
following DSMC simulations.


\subsection{Non uniform solutions}

In the present section we illustrate 
the results obtained simulating the system with the
heat bath (with recipe 2, i.e. equation~\eqref{recipe2}) by the
so called Direct Simulation Monte Carlo according to the
implementation described in~\cite{Baldassa}.

At every time step of length $\Delta t$ each particle 
is selected to collide with a
probability $p_c=\Delta t/\tau_c$ (where $\tau_c$ is an {\em a priori}
fixed mean free time established consistently with the mean free path
and mean squared velocity) and seeks its collision partner among the
other particles in a neighborhood of radius $r_B$, choosing it
randomly with a probability proportional to their relative velocities.
Moreover in this approximation the diameter $\sigma$ is no more
explicitly relevant, but it is directly related to the choices of $p_c$
and $r_B$ in a non trivial way: in fact the Bird algorithm allows the
particles to pass through each other, so that a rigorous diameter
cannot be defined or simply estimated as a function of $p_c$ and
$r_B$. In this section, to indicate the degree of damping, we give the
time $\tau_b=1/\beta$ instead of $\beta$. This is useful to
appreciate the ratio between the mean collision time $\tau_c$ and the
mean relaxation time due to the bath which indeed is $\tau_b$.

\begin{figure}[htbp]
\begin{center}
\includegraphics[clip=true,width=0.7\textwidth]{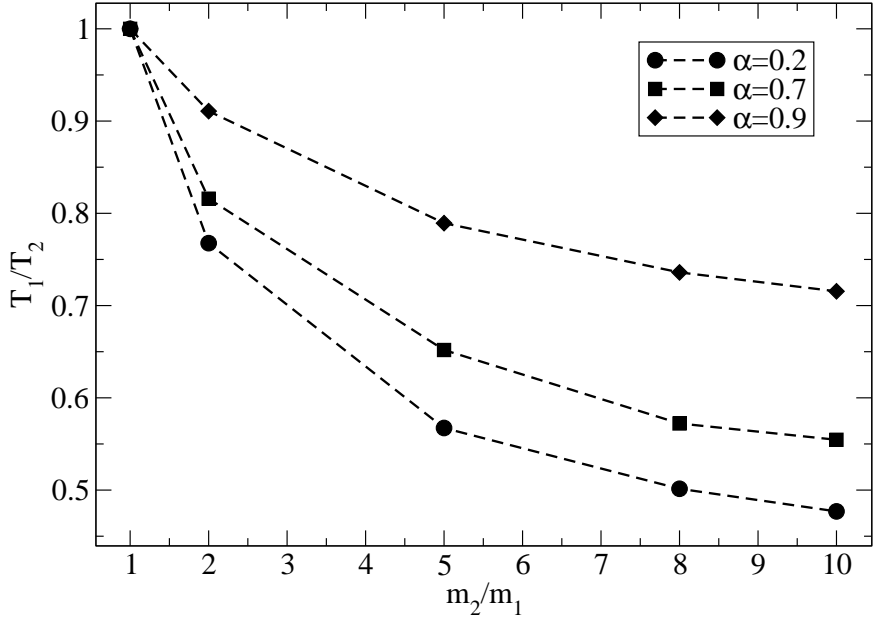}
\end{center}
\caption{
Granular temperature ratios $T_1/T_2$ vs. mass ratio $m_2/m1$ for a
binary mixture (DSMC simulation) with different values of $\alpha$,
$N_1=N_2=500$, $L^2=1000$, $T_b=1$, $\tau_b=10$, $\tau_c=0.16$, and
recipe $2$}
\label{rappvsmass}
\end{figure}

\begin{figure}[htbp]
\begin{center}
\includegraphics[clip=true,width=0.7\textwidth]{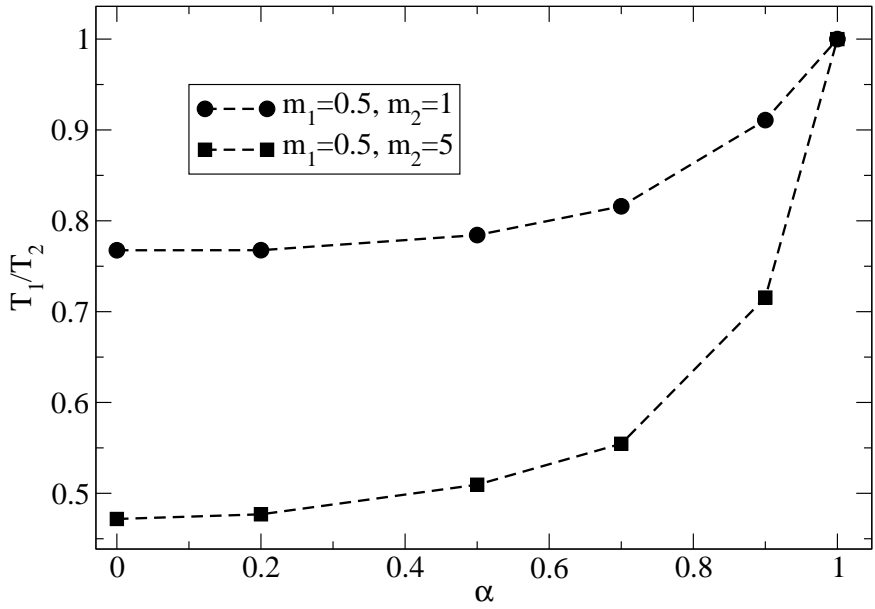}
\end{center}
\caption{
Granular temperature ratios $T_1/T_2$ vs. restitution coefficient $\alpha$ for a
binary mixture (DSMC simulation) with different values of $m_1/m_2$, $N_1=N_2=500$,
$L^2=1000$, $T_b=1$, $\tau_b=10$, $\tau_c=0.16$, and recipe 
$2$}
\label{rappvsalpha}
\end{figure}

In the present section we chose $N_1=N_2=500$ and $T_b=1$, and equal
restitution coefficients for all collisions and $\tau_c=0.16$. As
illustrated in figures~\ref{rappvsmass} and~\ref{rappvsalpha} the two
components display different granular temperatures in agreement with
the analytical predictions of the homogeneous Boltzmann equations.  We
also comment that the ratio $T_1/T_2$ is very sensitive to the mass
ratio and much less to differences in the restitution
coefficients and concentrations. This is in agreement with
experimental observations~\cite{feitosa}.

\begin{figure}[htbp]
\begin{center}
\includegraphics[clip=true,width=0.7\textwidth]{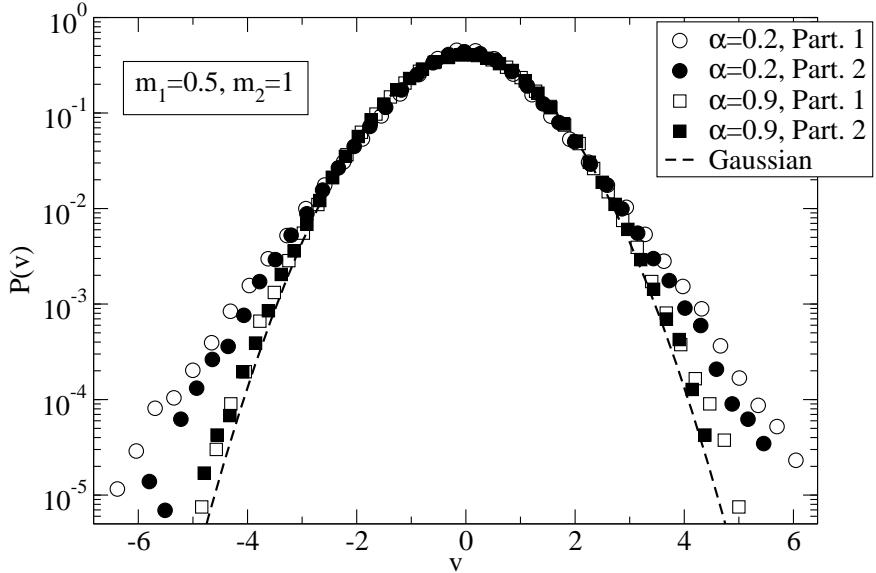}
\end{center}
\caption{
Rescaled (to have variance $1$) velocity distributions $P(v)$ vs. $v$ in the numerical experiment with
the thermal bath (DSMC simulation), for binary mixtures of particles with mass $m_1=0.5$
and $m_2=1$, with $N_1=N_2=500$,
$L^2=1000$, $T_b=1$, $\tau_b=10$,  $\tau_c=0.16$, recipe  $2$, and
different values of $\alpha$}
\label{bulk_zeta2}
\end{figure}

\begin{figure}[htbp]
\begin{center}
\includegraphics[clip=true,width=0.7\textwidth]{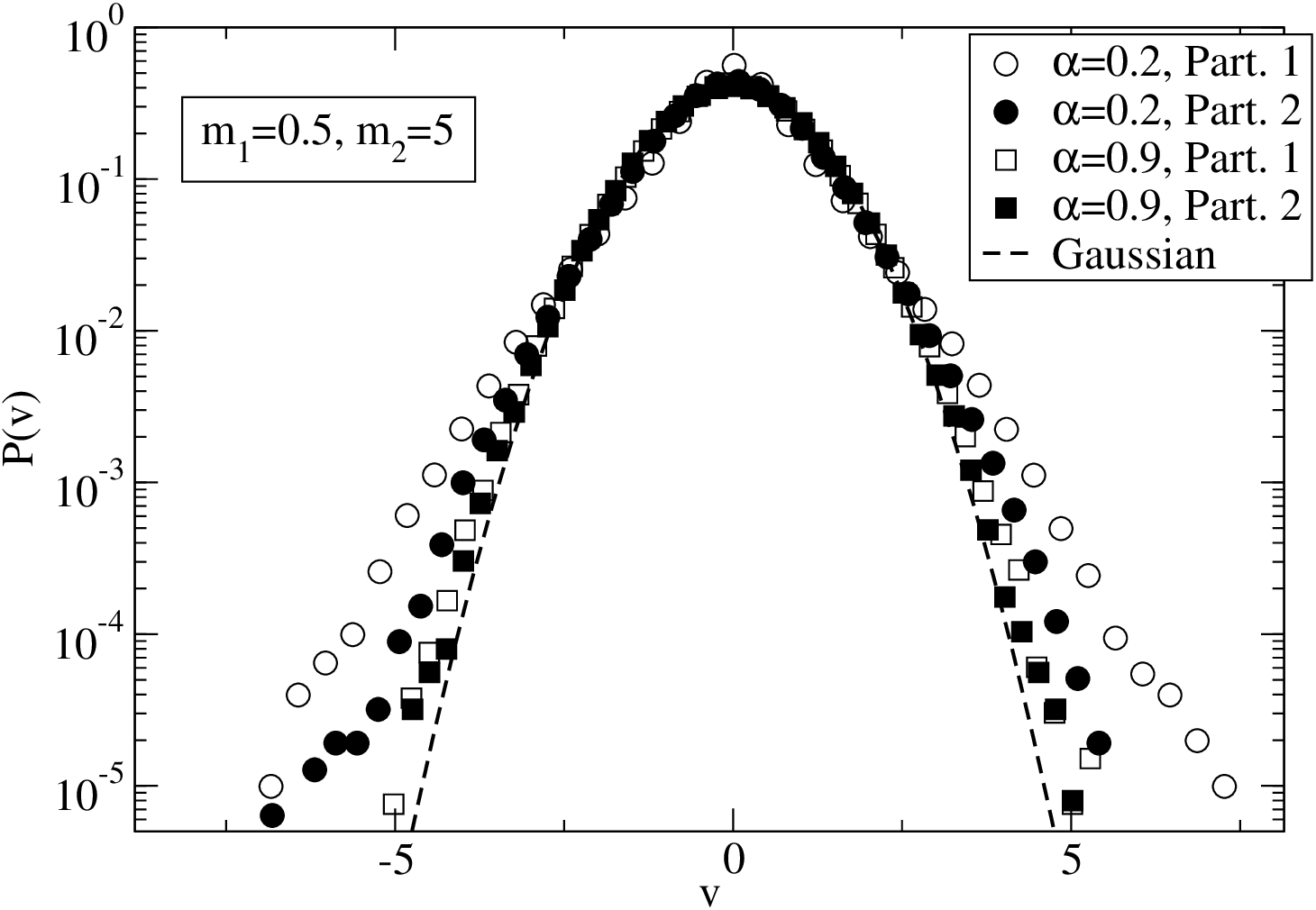}
\end{center}
\caption{
Rescaled (to have variance $1$) velocity distributions $P(v)$ vs. $v$
in the numerical experiment (DSMC simulation) with the thermal bath,
for binary mixtures of particles with mass $m_1=0.5$ and $m_2=5$, with
$N_1=N_2=500$, $L^2=1000$, $T_b=1$, $\tau_b=10$, $\tau_c=0.16$, recipe
$2$, and different values of $\alpha$}
\label{bulk_zeta10}
\end{figure}

At a finer level of description we consider the rescaled velocity pdf
for different values of the inelasticity parameter and different mass
ratios $m_2/m_1=2$. One sees that not only the deviations from the
Gaussian shape become more and more pronounced as we increase the
inelasticity parameter $(1-\alpha)$, but also the shape of the two
distributions differ appreciably in the tails even after velocity
rescaling to make the two pdf's have the same variance. Such a
property is similar to the one already reported in~\cite{ioepuglio}.
One can also observe that the rescaled pdf of the lighter species has
slightly broader tails. 
The mass ratio also controls the deviations from the Gaussianity of
the velocity pdf's. Whereas it is known that the departure from
the Maxwell-Boltzmann statistics is triggered by the inelasticity of
the collisions, the larger the inelasticity the stronger the
deviation, to the best of our knowledge 
this is the first numerical evidence of the
phenomenon, predicted within a Maxwell model in ref.~\cite{ioepuglio}.
Comparing figs. \ref{bulk_zeta2} and \ref{bulk_zeta10}
one sees that the mass asymmetry enhances the non gaussianity
of the pdf. 

\begin{figure}[htbp]
\begin{center}
\includegraphics[clip=true,width=0.7\textwidth]{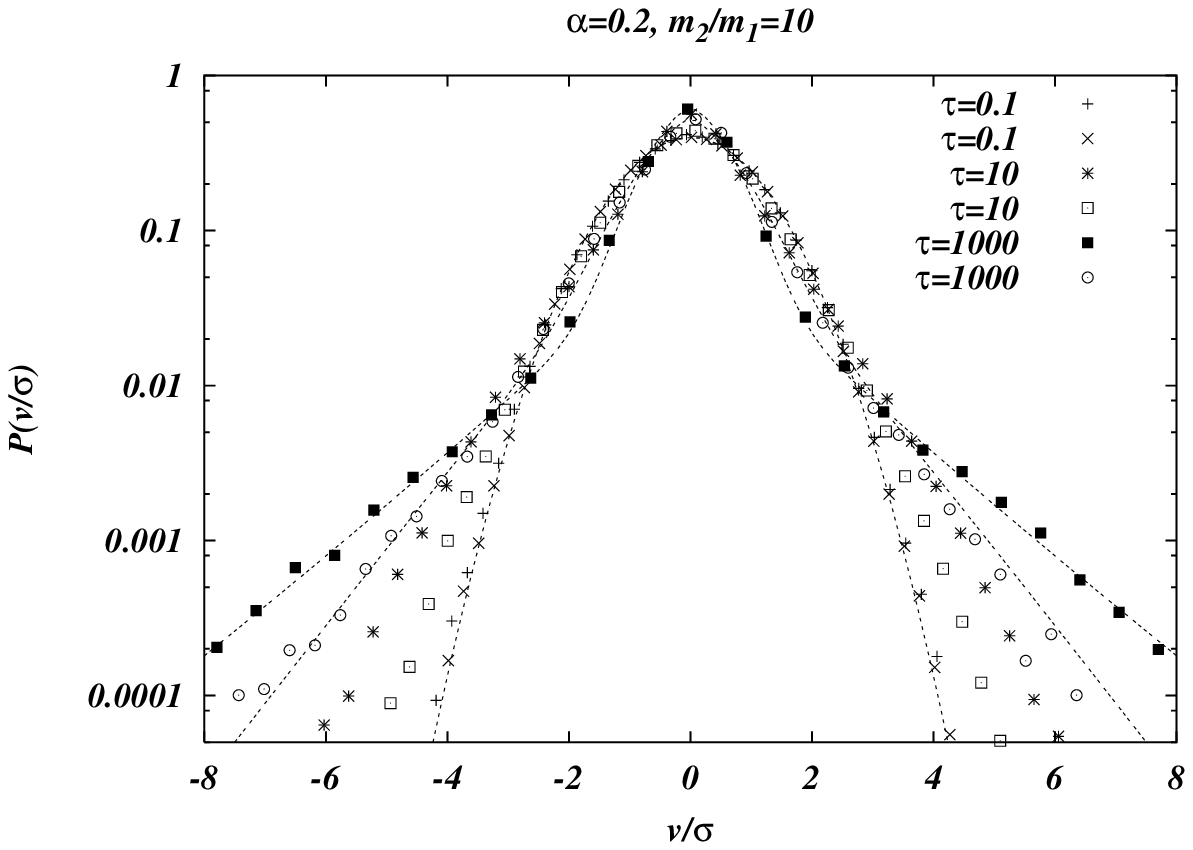}
\end{center}
\caption{
Rescaled (to have variance $1$) velocity distributions $P(v)$ vs. $v$
in the numerical experiment (DSMC simulation) with the thermal bath,
for binary mixtures of particles with mass $m_1=0.5$ and $m_2=5$, with
$N_1=N_2=500$, $L^2=1000$, $\alpha=0.2$, $T_b=1$, different values for
$\tau_b$, $\tau_c=0.16$, with recipe $2$, and different values of
$\alpha$}
\label{distrib_varitau}
\end{figure}

\begin{figure}[htbp]
\begin{center}
\includegraphics[clip=true,width=0.7\textwidth]{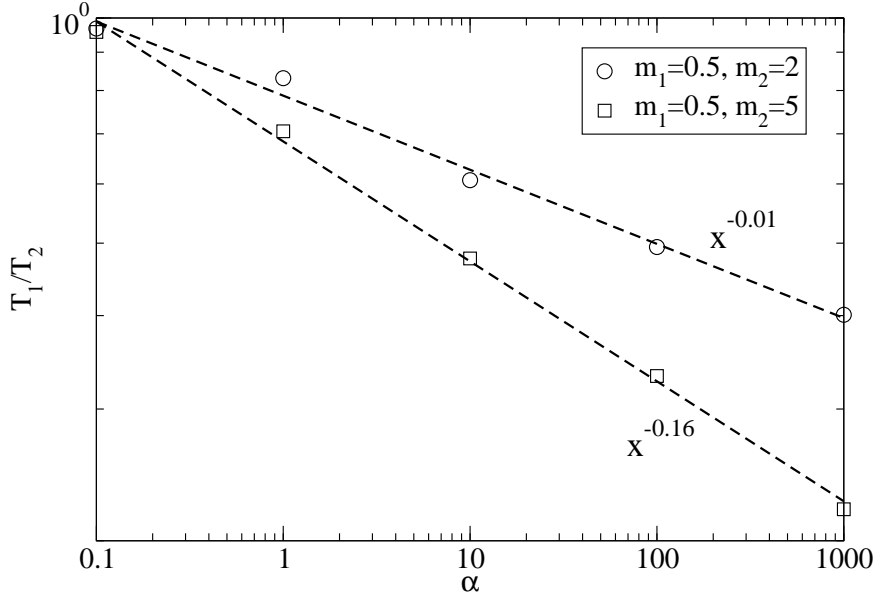}
\end{center}
\caption{
Temperature ratios $T_1/T_2$ vs. the viscosity time $\tau_b$ in the
numerical experiment (DSMC simulation) with the thermal bath for a
binary mixture of particles with mass $m_1=0.5$ and $m_2=5$, with
$N_1=N_2=500$, $L^2=1000$, $\alpha=0.2$, $T_b=1$, different mass
ratios, $\tau_c=0.16$, with recipe $2$, and different values of
$\alpha$}
\label{t1t2vstaub}
\end{figure}

We have also studied the limits of low and high $\tau_b$, in
figure~\ref{distrib_varitau}, to show how the velocity distributions
change. For values of the characteristic time of the heat bath,
$\tau_b$, comparable with the collision time, $\tau_c$, the dynamics
is essentially controlled by the stochastic acceleration term. This
fact renders the two partial temperatures very close and makes the
velocity distributions nearly Maxwellian.  As $\tau_b$ increases we
have observed that the energy dissipation due to the inelasticity
makes the temperatures of the two species different.  Moreover, the
temperature ratio displays power law decreasing trend as a function of
$\tau_b$ whose strength depends on the mass ratio (see figure~\ref{t1t2vstaub}).

In order to obtain information about the spatial structure of the 
mixture we have performed an analysis of the following correlation
function, already introduced in the context of granular media
by~\cite{puglisi1,puglisi2,procaccia}:

\begin{equation}
C_{\alpha\beta}(r)=
\frac{1}{N(N-1)}\sum_{i\neq j} \Theta(r-|\mathbf{x}_i^{\alpha}-\mathbf{x}_j
^{\beta}|)
\end{equation}

\begin{figure}[htbp]
\begin{center}
\includegraphics[clip=true,width=0.7\textwidth]{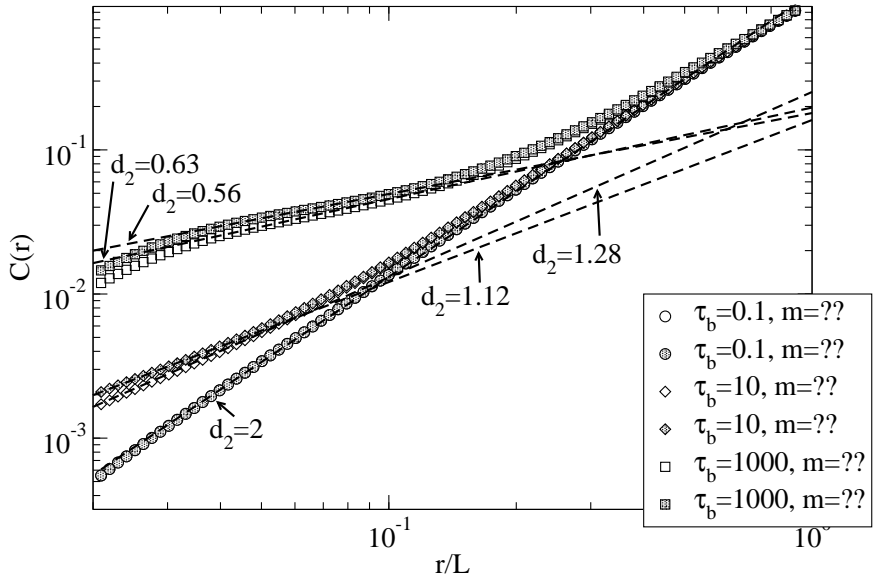}
\end{center}
\caption{
Correlation function used to compute the correlation dimension $d_2$,
$C(r)$ vs. $r$ (see text for definition), for a binary mixture (DSMC
simulation) with $m_1=0.5$, $m_2=5$, $\tau_c=0.16$, $T_b=1$,
$N_1=N_2=500$, $L^2=1000$, $\alpha=0.2$, with different values for
$\tau_b$ }
\label{structure}
\end{figure}

For a spatially homogeneous system 
we expect that $C_{\alpha\beta}(r)\approx r^{d_2}$ with
$d_2=d$ the dimension of the embedding space. However, in the cases we
have studied, we found that $d_2<d$ 
(see fig~\ref{structure}), a signature that the
system tends to clusterize. The larger the value of $\tau_b$ the
stronger the deviation.

\section{System with gravity and heated from below}

We turn, now, to illustrate the results relative to an inelastic mixture
subject to gravity and confined to a vertical plane of dimensions $L_x
\times L_y$. In the horizontal direction $x$ we assumed periodic
boundary conditions. Vertically the particles are confined by
walls. Energy is supplied by the bottom wall vibrating stochastically
or periodically according to the method employed in \cite{Baldassa}
for a one component system.

The vibration can have either a periodic character (as in
ref.~\cite{kudrolli}) or a stochastic behavior with thermal properties
(as in~\cite{japan}). In the periodic case, the wall oscillates
vertically with the law $Y_w=A_w \sin(\omega_w t)$ and the particles
collide with it as with a body of infinite mass with restitution
coefficient $\alpha_w$, so that the vertical component of their
velocity after the collision is $v_y'=-\alpha_wv_y+(1+\alpha_w)V_w$
where $V_w=A_w \omega_w
\cos(\omega_w t)$ is the velocity of the vibrating wall. In the
stochastic case we assume that the vibration amplitude is negligible
and that the particle colliding with the wall have, after the
collision, new random velocity components $v_x \in (-\infty,+\infty)$
and $v_y\in (0,+\infty)$ with the following probability distributions:

\begin{equation} 
P(v_y)=\frac{v_y}{T_w}exp(-\frac{v_y^2}{2 T_w}) 
\end{equation}

\begin {equation} 
P(v_x)=\frac{1}{\sqrt{2 \pi T_w}}exp(-\frac{v_x^2}{2 T_w}) 
\end{equation}

In this model we assume that the particle do not feel any external
(environmental) friction. In all simulation we have chosen
$N_1=N_2=100$, $g=1$, $m_1=m_2/2=1/4$, $L_x=L_y=L=\sqrt{N}$.

The Boltzmann equations for the partial distribution functions 
$f_{\kappa}({\bf r,v};t)$ with $\kappa=1,2$ read

\begin{equation}
\left( \frac{\partial}{\partial t} + {\bf v} \cdot \nabla_r+ g_i
\frac{\partial}{\partial v_i} \right) 
f_{\kappa}({\bf r},{\bf v},t)=\sum_{\beta}J_{\kappa \beta}
(f_{\kappa},f_{\beta}).
\label{boltzeq} 
\end{equation}

The simulation results show that the tendency of the grains to form
clusters is enhanced by such a choice of driving mechanism with
respect to the homogeneous heat bath of the previous section. In the
latter the noise acting uniformly was more effective in breaking the
clusters. In addition the gravitational force tends to group the
particles in the lower portion of the container
for not too large driving frequencies \cite{Baldassa}. 

\begin{figure}[htbp]
\begin{center}
\includegraphics[clip=true,width=0.7\textwidth]{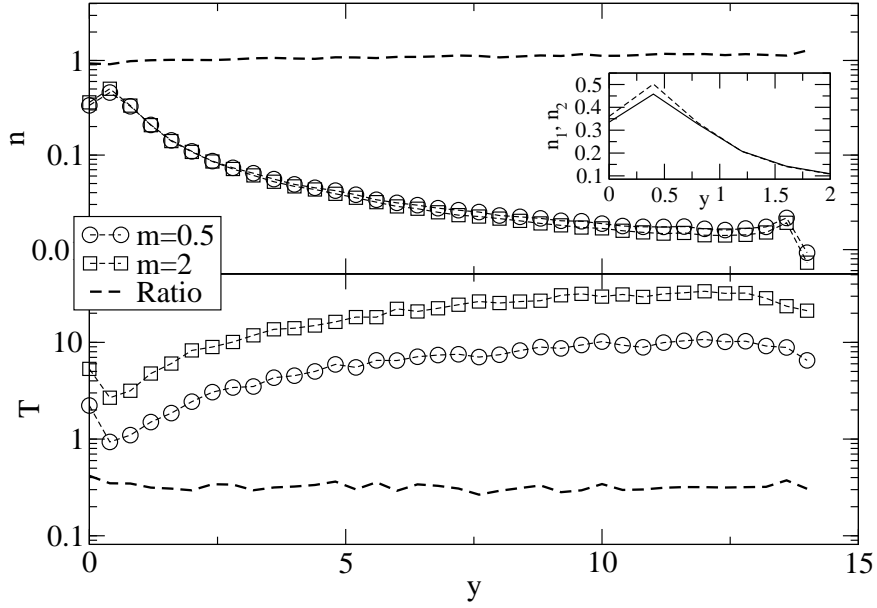}
\end{center}
\caption{
Profiles for density and granular temperature $n(y)$, $T(y)$ vs. $y$
for a binary mixture with $m_1=0.5$, $m_2=2$, on an inclined plane
with a thermal bottom wall (DSMC simulation) with $T_w=20$,
$N_1=N_2=100$, $L^2=200$, $\alpha=0.6$, $\alpha_w=1$, $g=1$; the
dashed lines indicate the ratios for both profiles. }
\label{gravity_profiles}
\end{figure}

\begin{figure}[htbp]
\begin{center}
\includegraphics[clip=true,width=0.7\textwidth]{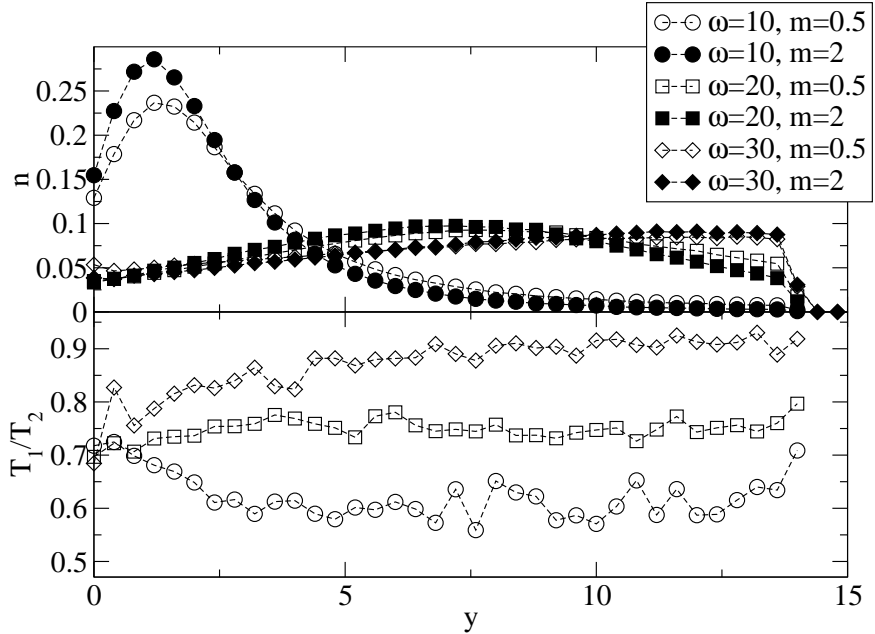}
\end{center}
\caption{
Profiles for density $n(y)$ vs. $y$ and ratios between granular
temperatures of the two species $T_1(y)/T_2(y)$ vs. $y$, for a binary
mixture (DSMC simulation) with $m_1=0.5$ and $m_2=2$, on an inclined
plane with an harmonic oscillating bottom wall with $A=1$, different
values of $\omega$, $N_1=N_2=100$, $L^2=200$, $\alpha=0.9$,
$\alpha_w=\alpha$ $g=1$.}
\label{gravitysin_profiles}
\end{figure}

Fig~\ref{gravity_profiles} illustrates the partial density profiles and
granular temperature profiles in the presence of a thermal wall of
intensity $T_p=20$. The density profiles differ slightly near the
bottom wall where both present a maximum (see inset). The temperature
profiles are different, putting again in evidence a strong lack of
equipartition in the system. Interestingly, the temperature ratio is
almost constant along the vertical direction, notwithstanding 
the partial profiles are
non constant. We have also performed
numerical simulations with a harmonically vibrating wall: as the maximum
velocity of the vibrating wall (which, for $A=1$, is equal to
$\omega$) increases, the position of the density maximum raises
indicating that gravity becomes less and less relevant. As far as the
partial temperatures are concerned figure~(\ref{gravitysin_profiles})
illustrates the corresponding situation.

One sees that the temperatures (see inset) next to the wall attain
their largest value, then drop to increase again. This
indicates that the region far from the bottom is hotter 
because of the
lower density of the gas and of the lower collision rate.
As the wall
temperature increases the temperature becomes more homogeneous far from the
bottom due to the major homogeneity in density. We also notice a small
segregation effect.

\begin{figure}[htbp]
\begin{center}
\includegraphics[clip=true,width=0.7\textwidth]{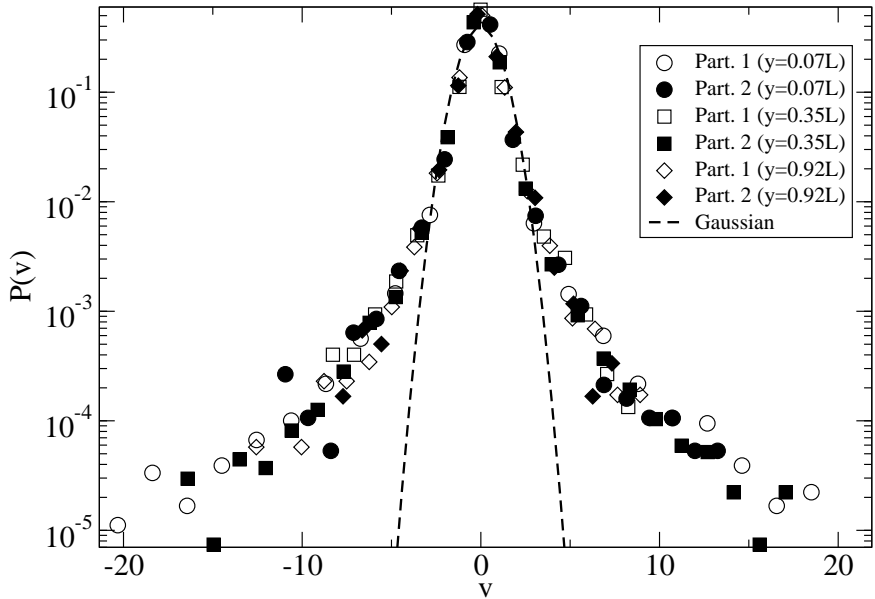}
\end{center}
\caption{
Rescaled distributions (to have variance $1$) of velocity $P(v)$
vs. $v$ for particles taken in stripes at different heights, for a
binary mixture (DSMC simulation) with $m_1=0.5$, $m_2=2$, on an
inclined plane with a thermal bottom wall with $T_w=20$,
$N_1=N_2=100$, $L^2=200$, $\alpha=0.6$, $g=1$.}
\label{gravity_distrib}
\end{figure}

Finally, we present the velocity distributions measured in the system
with a thermal wall.  Since the temperature depends on the vertical
coordinate $y$ we have computed the velocity pdf at different
heights. The various pdf are plotted in figure~(\ref{gravity_distrib})
after a suitable rescaling. We observe that the distributions deviate
appreciably from a Gaussian shape and display overpopulated high
energy tails.

Interestingly the value obtained by our simulation
for the temperature ratio, having chosen
$\frac{m_1}{m_2}=0.3$ and a single restitution coefficient
$\alpha=0.93$, $\frac{T_1}{T_2}\approx 0.75$
is not too far from the value obtained
experimentally by Feitosa and Menon~\cite{feitosa} which is
$0.66 \pm 0.06$. One should recall that in our ``setup'' only
the lower wall supplies energy to the system, whereas in the
experiment both walls vibrate.



\section{Conclusions}

To summarize we have studied the steady state properties of a granular
mixture subject to two different classes of external drive.

In the case of an heat bath acting homogeneously on the grains, we
have first obtained the temperatures of each species by employing an
approximate analytic method. We have then
solved numerically the equations by allowing the density 
and the temperatures to be
spatially varying and analyzed the spatial correlations and the
velocity distributions.

Finally, we turned to investigate the properties of the mixture in the
presence of gravity and inhomogeneous drive.

The present results not only confirm the predictions 
of different temperatures
for the two species for all types of drive, but also show the
existence of different shape functions for the velocity distributions
and of overpopulated high energy tails in agreement with the findings
based on Maxwell models \cite{ioepuglio}.

In order to obtain a direct comparison with experiments we employed
parameters comparable to those utilized in the experimental work of
ref.~\cite{feitosa} and found temperature ratios not too far from
the  ones there reported.  
In spite of this success, the agreement seems to be limited
to the average properties, while it is clear from the observation of
the density profiles computed numerically that a better treatment of the
excluded volume effect is necessary in order to obtain a more
realistic description of the system.  Finally, in view of the approximations
inherent to the Boltzmann approach, we have not tried to include
neither the effect of rotations of the grains nor the
friction with the lateral
walls which might be both relevant \cite{Paolotti}.

\begin{acknowledgments}
This work was supported by Ministero dell'Istruzione,
dell'Universit\`a e della Ricerca, Cofin 2001
Prot. 200102384. A. P. acknowledges support from the INFM Center for
Statistical Mechanics and Complexity (SMC).
\end{acknowledgments}

\end{document}